\documentclass[groupedaddress]{revtex4}

\usepackage{graphicx}

\begin{document}

\title{Virtual Reality Visualization by CAVE\\ with VFIVE and VTK}

\author{Nobuaki Ohno}
\author{Akira Kageyama}
\author{Kanya Kusano}
\affiliation{Earth Simulator Center, \\
Japan Agency for Marine-Earth Science and Technology\\
3173-25 Showa-machi, Kanazawa-ku, Yokohama, 236-0001 Japan}


\begin{abstract}
The CAVE-type virtual reality (VR) system was introduced for scientific visualization of large scale data 
in the plasma simulation community about a decade ago. 
Since then, we have been developing a VR visualization software, VFIVE, for general CAVE systems.
Recently, we have integrated an open source visualization library, the Visualization Toolkit (VTK), into VFIVE.  
Various visualization methods of VTK can be incorporated and used interactively in VFIVE. 
\end{abstract}

\maketitle

\section{Introduction}
Simulation science is supported by two key technologies; the computation and visualization. 
While the computer technology still keeps the exponential growth, 
its counterpart, the visualization technology, does not.
Therefore the gap between the two key technologies is exponentially growing. 
It is becoming harder for simulation scientists to understand the output data by means of 
the traditional visualization technology based on graphic workstations or PCs. 
We believe that the modern virtual reality (VR) technology provides the solution to this technological imbalance. 
Among various VR systems developed to date, the CAVE-type VR system is especially suited   
to our purpose, i.e., the visualization of complicated 3-D simulation data. 
\begin{figure}[h]
\centering
\includegraphics[width=4.5cm]{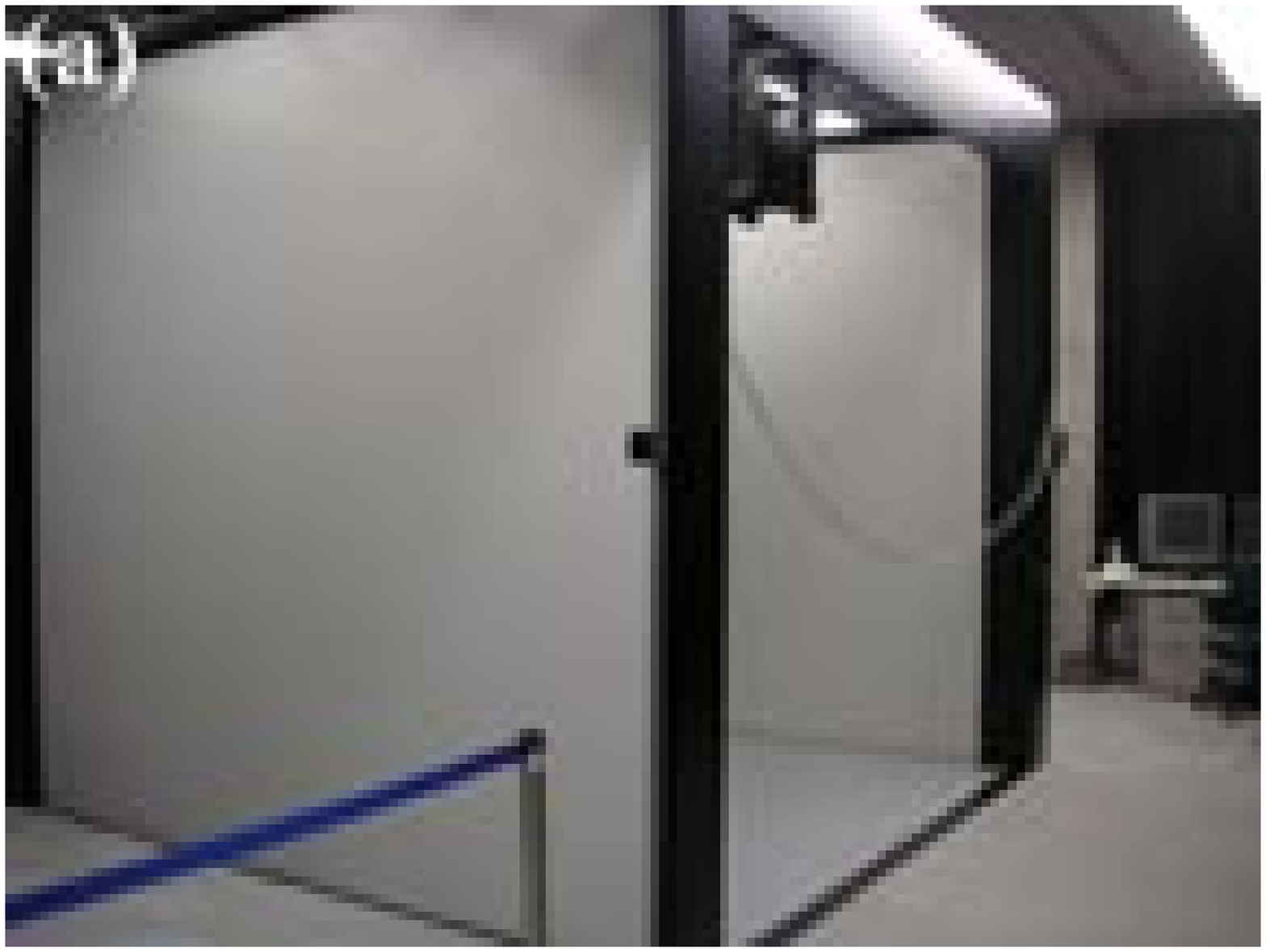}
\includegraphics[width=3.8cm]{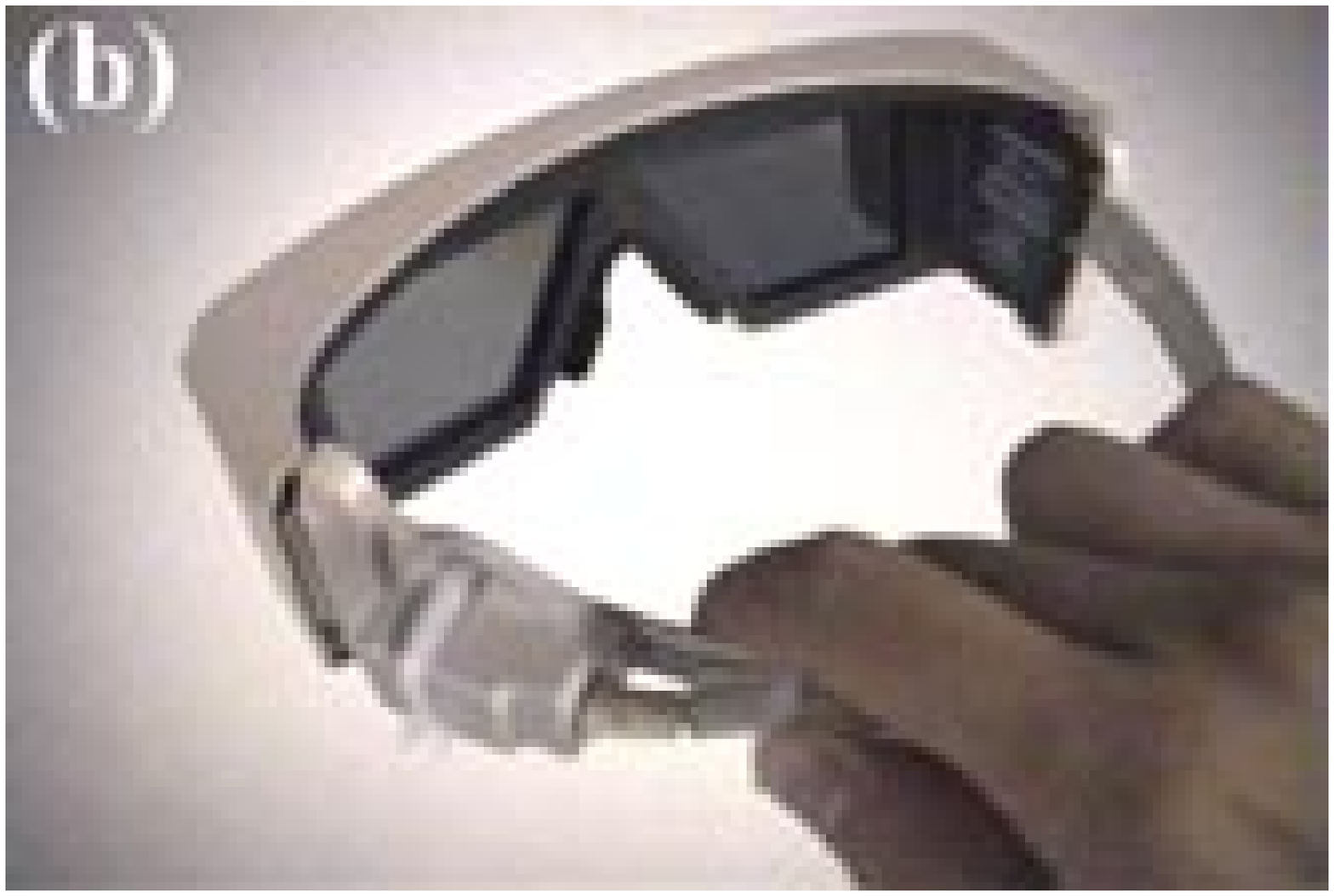}
\includegraphics[width=3.8cm]{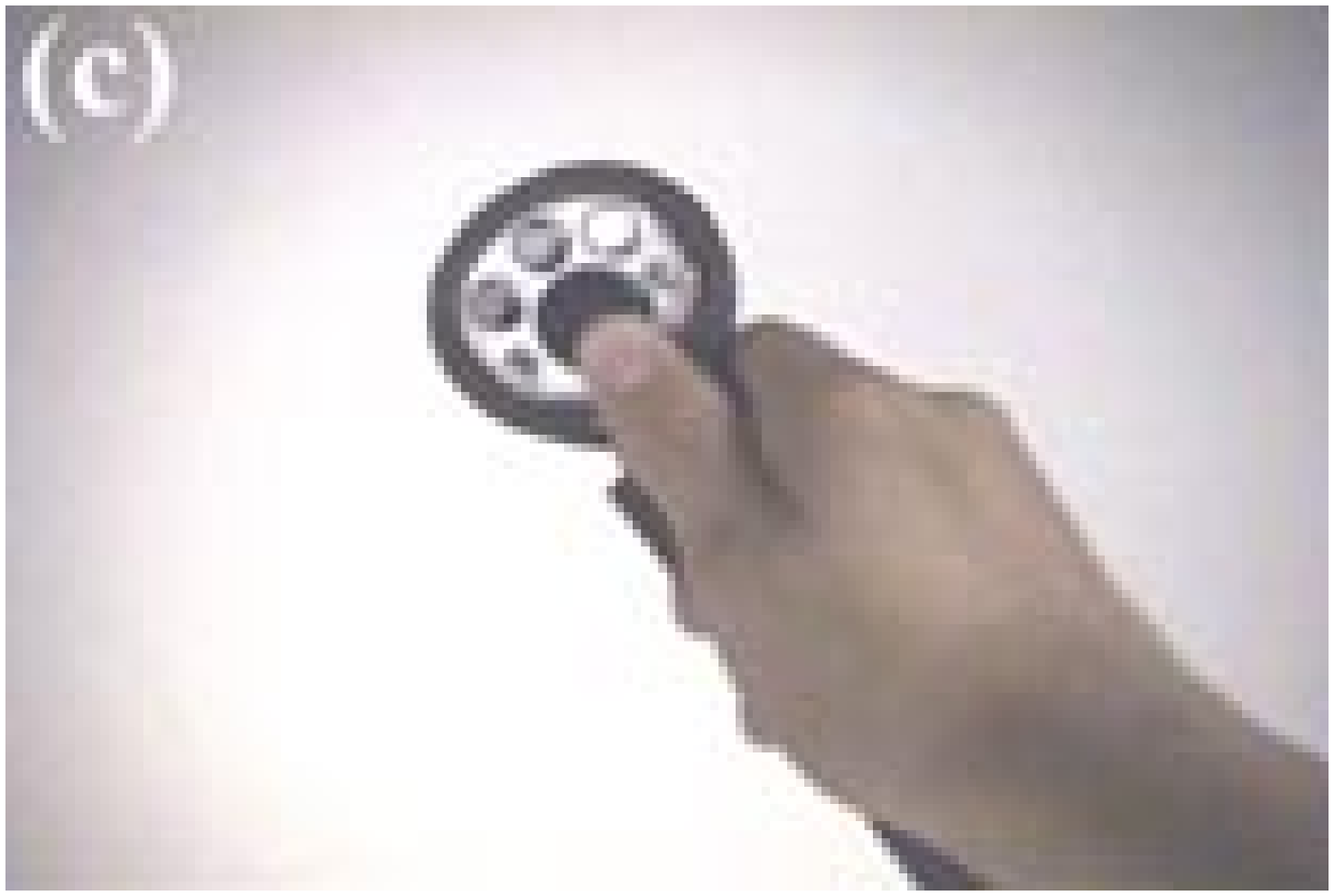}
\caption{({\it a}) Our CAVE system, ``BRAVE". ({\it b}) Liquid crystal shutter glasses. 
({\it c}) A portable controller, ``wand".}
\label{fig:cave}
\end{figure}
The CAVE is a room-sized, cubic-shaped VR system developed at University of Illinois, Chicago (\cite{cruz}). 
We installed a CAVE system ``BRAVE" which is shown in Fig.~\ref{fig:cave}({\it a}), in our facility in 2003. 
Stereo Images are projected onto its $3$m$\times$$3$m screens of the walls and the floor.
The viewer stands in the CAVE's room wearing a liquid crystal shutter glasses shown in Fig.~\ref{fig:cave}({\it b}).  
A magnetic sensor is attached to the glasses so that the projected images are automatically adjusted
to the viewer's position in real time. 
The stereo images for the four screens ($3$ walls $+$ $1$ floor) are generated by four DLP projectors.  
Since they are projected seamlessly on their borders, everything looks natural from the viewer inside the CAVE and 
feels a deep immersion in the VR world. 
The viewer interacts with the simulation data through a portable controller, ``wand", shown in Fig.~\ref{fig:cave}({\it c}). 
Another magnetic sensor is attached to the wand to detect its position and direction.

\section{VFIVE and VTK}
VFIVE is our original virtual reality visualization software for CAVE systems (\cite{kage99}, 2000). 
The major visualization methods of VFIVE are listed in Table~\ref{tbl:vismeth}. 
VFIVE is a fully interactive visualization tool; we can control the isosurface level 
by the vertical motion of the wand in the CAVE room; 
we can start new tracer particles from the tip of the wand by pressing a wand button; 
and a spotlight is emitted by the wand and thousands of tracer particles are flying in the cone-shaped light.
In the original version, VFIVE has only basic visualization methods such as Isosurface and Field Lines. 
In order to add a new visualization method to VFIVE, one had to write a subset of computer graphics (CG) software 
with OpenGL and CAVELib. 
In contrast to usual CG programs, a CAVE application programmer does not have to write the projection part of the software 
since the CAVELib automatically takes care about that part. 
The programmer writes the {\it display process} part and the {\it application computation process} part.
The former is to draw images on the screens and the latter is for calculations, 
e.g., the orbits of tracer particles.
Therefore, simulation researchers who want a new visualization method in VFIVE for their own analysis need 
a fairly good knowledge and experience of programming with OpenGL and CAVELib but 
it is clearly overwhelming for most simulation researchers. 
This has been a hurdle for VFIVE to become 
an actively used research tool in the simulation science.
To resolve this problem, we have made a major upgrade of VFIVE by incorporating a general visualization library, 
the Visualization Toolkit (VTK).

\begin{table}
\caption{Major visualization methods of VFIVE}
\label{tbl:vismeth}
\begin{tabular}{ll}
\hline
For Scalar data & For Vector data \\
\hline
{\it Isosurface}              & {\it Field Lines} \\
{\it Color Contour}           & {\it Particle Tracer}\\
{\it Local Slice Plane}       & {\it Vector Allows} \\
{\it Contour Lines} (VTK)     & {\it Spotlighted Particles} \\
                              & {\it Stream Surfaces} (VTK)\\
                              & {\it Stream Tubes} (VTK)\\
\hline
\end{tabular}
\end{table}
VTK (\cite{vtk}) is an open source, freely available visualization software 
including many sophisticated visualization methods for scalar, vector, and tensor fields. 
They are implemented as a C$++$ class library. 
The usage is very simple: Choose the modules in the library and combine them by the so called ``VTK pipeline",
then you will get a full set of high performance visualization program without a detailed knowledge of the visualization algorithm 
and OpenGL. 
It is an attractive idea for CAVE users to integrate VTK into CAVE applications (\cite{leigh}).

\section{Integrating VTK into VFIVE}
Many visualization methods of VTK are based on the polygon-based rendering.
This means that each visualization module converts the input field data to a set of numerous numbers of discrete polygons. 
The polygonal data consists of the vertex positions, vertex colors, and the normal vectors.
In order to combine VTK with VFIVE, we need to show the VTK polygons in the CAVE's 3-D virtual space.
However, to use VTK with the CAVELib is not easy due to the design mismatch between these two libraries. 

To use VTK in the CAVE, 
Rajlich (1998) created a software ``vtkActorToPF" to pull out polygonal data of VTK and 
render them using IRIS Performer (or OpenGL Performer). 
Hall (1999) created ``vtk2CAVE", in which the {\it display processes} access the vtkActor
copied in shared memory and draw its polygonal data by OpenGL. 
Shamonin (2002) created a library ``VtkCave", 
which makes it possible to create a CAVE application with VTK. 

We took another approach to combine VTK into our CAVE program, VFIVE.
Basically, we have followed the idea of ``vtkActorToPF", but we did not want to use IRIS Performer 
because VFIVE is written directly on OpenGL, without IRIS Performer.
Therefore, we have revised VFIVE in such a way that the {\it application computation process} 
can make the polygonal data by using VTK and send them via hard disk drive to the {\it display processes}, 
and the {\it display processes} can receive the output polygonal data of VTK and draw them by OpenGL. 
In this way we have succeeded to integrate important polygon-based visualization methods of VTK,
for example, Contour Lines and Stream Surfaces.
It is easy to integrate other VTK modules into VFIVE. 

\begin{figure}[htb]
\begin{center}
\includegraphics[height=4.1cm]{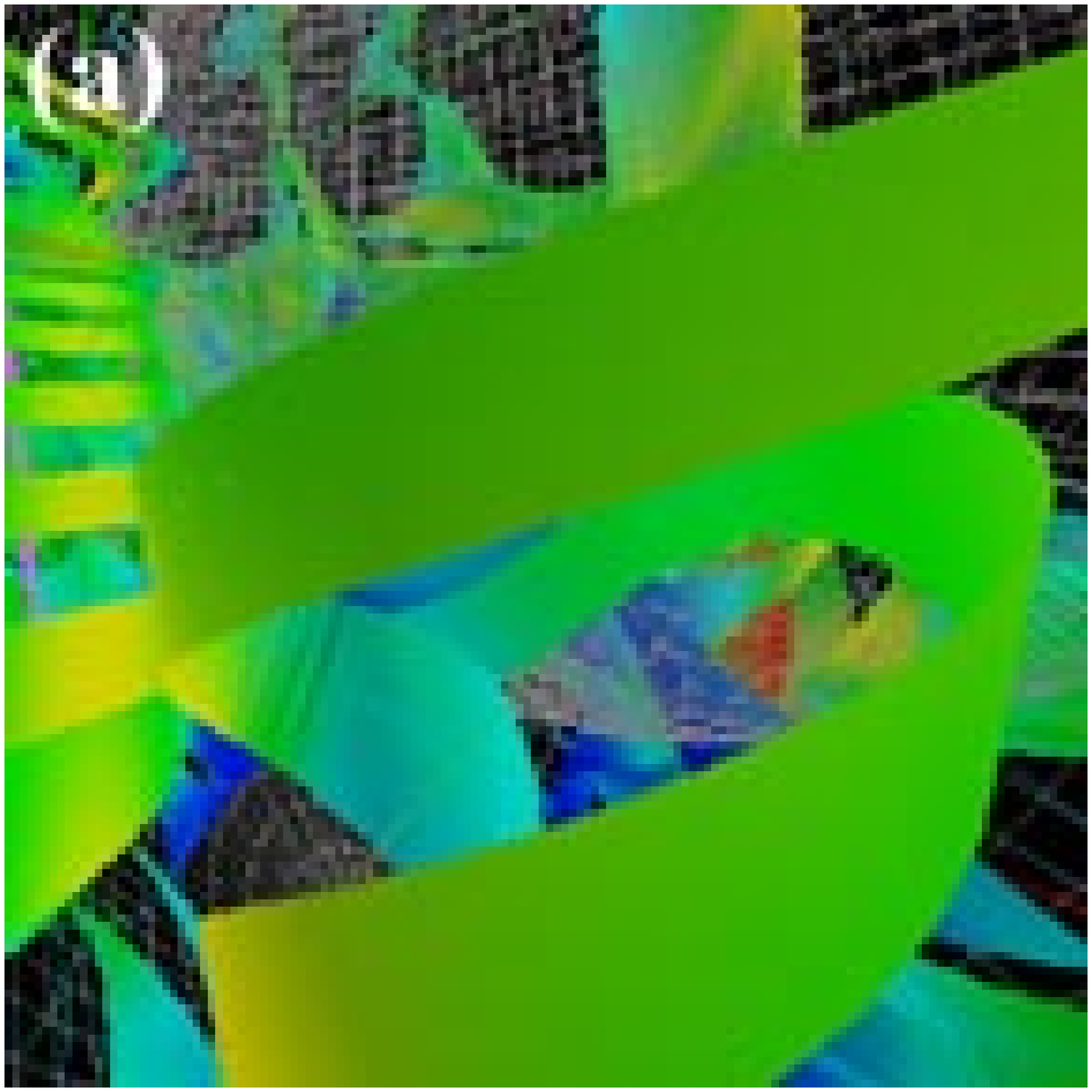}
\includegraphics[height=4.1cm]{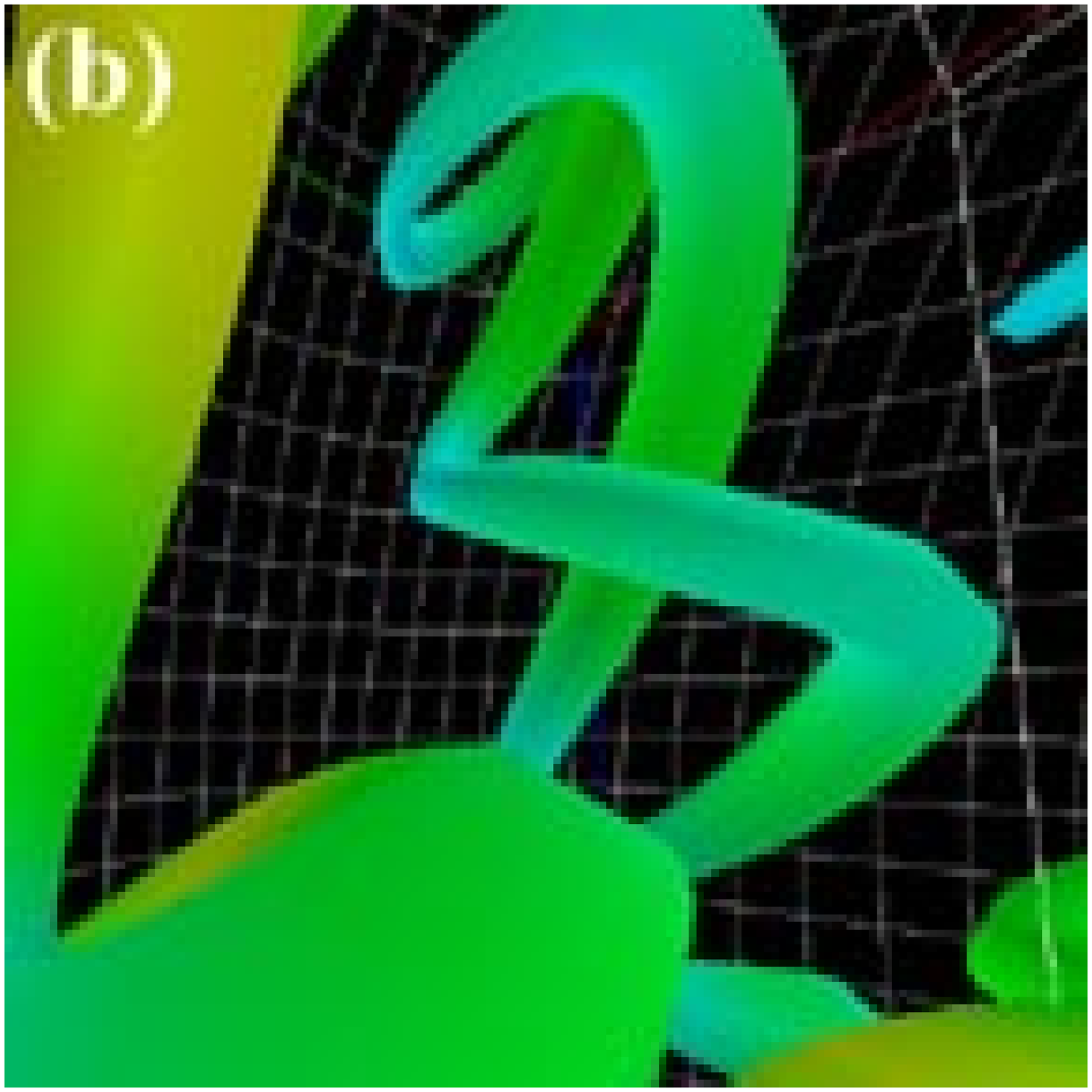}
\includegraphics[height=4.1cm]{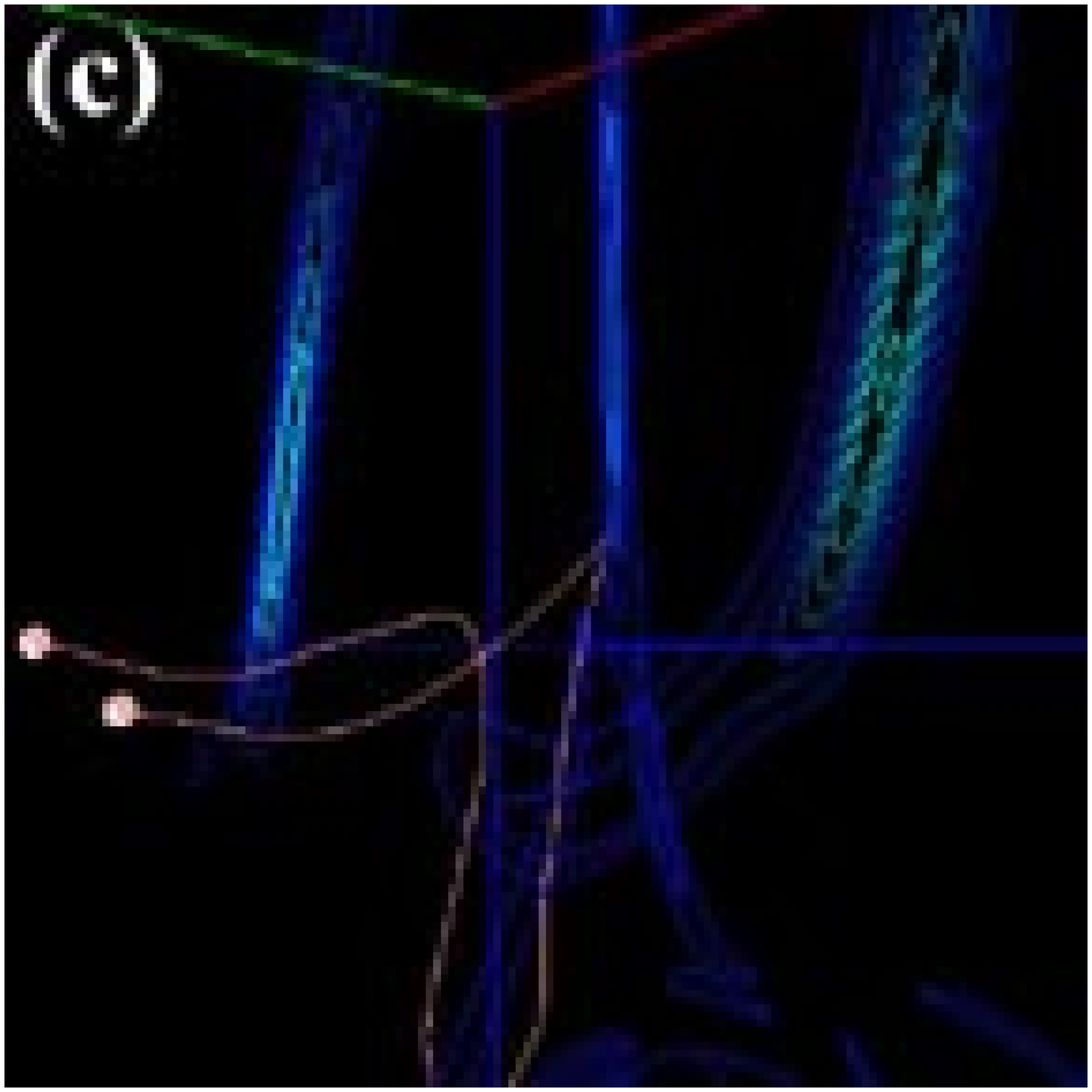}
\end{center}
\caption{
({\it a}) Stream Surfaces of the flow velocity of a geodynamo simulation. 
({\it b}) Stream Tubes of the flow velocity of a geodynamo simulation.
({\it c}) Contour Lines of the magnitude of the electric current and Field Lines of the flow velocity of a solar coronal simulation.
}
\label{fig:vrvtk}
\end{figure}

\begin{figure}[htb]
\begin{center}
\includegraphics[height=4.67cm]{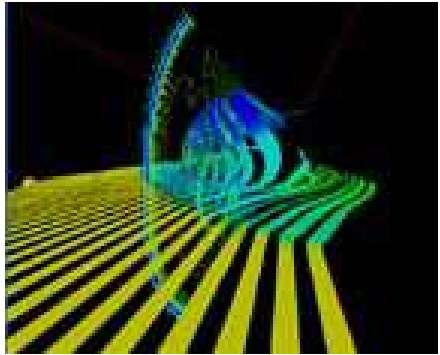}
\end{center}
\caption{
Stream Surfaces of the flow velocity and Contour Lines of the pressure of the Earth's magnetosphere simulation. 
}
\label{fig:mag}
\end{figure}

Here we show the power of revised VFIVE enhanced by VTK in the applications to  
MHD simulations.
Figure \ref{fig:vrvtk}({\it a}) shows a VTK visualization of a geodynamo simulation data by Stream Surfaces. 
This method visualizes a 3-D vector field by
ribbon-like surfaces created by multiple field lines started from a line segment. 
The colors of the ribbons indicate the magnitudes of the vector field. 
Figure \ref{fig:vrvtk}({\it b}) shows another VTK visualization, Stream Tubes. 
This method also visualizes a 3-D vector field by a field line wrapped by a tube by which  
more information than a simple 1-D field line can be conveyed. 
The colors of the tube indicate the magnitudes of the vector field and its radius depends on the vector magnitude.
Figure \ref{fig:vrvtk}({\it c}) shows Contour Lines of the electric current magnitude of a solar coronal simulation. 
This is also a VTK visualization method. 
Figure \ref{fig:mag} shows a mixed application of the two VTK visualization methods, 
Stream Surfaces and Contour Lines. 

\section{Conclusions}
We have integrated VTK into our virtual reality software VFIVE for 3-D, interactive visualization in the CAVE system.
The purpose of this improvement is to reduce the hurdle to add a new visualization function into VFIVE. 
VTK was chosen because of its rich and sophisticated visualization methods.

We believe that the development of an easy-to-use and easy-to-extend software for VR visualization in the CAVE, 
like our VFIVE, is important for simulation science in future.

\begin{acknowledgments}
We would like to thank Prof. T. Ogino for providing us 
the simulation data of the Earth's magnetosphere. 
\end{acknowledgments}

\end{document}